\documentclass[aps,prl,twocolumn,showpacs,superscriptaddress,unsortedaddress]{revtex4}
\usepackage{ae,aecompl}
\usepackage{graphicx}
\usepackage{psfrag}
\usepackage{amssymb}

\begin{document}
\title{Mechanism and Lineshapes of Sub-Fourier Resonances}
\author{Hans Lignier}
\author{Jean-Claude Garreau}
\author{Pascal Szriftgiser}
\affiliation{Laboratoire de Physique des Lasers, Atomes et Mol\'ecules, UMR CNRS 8523,
Universit\'e des Sciences et Technologies de Lille, F-59655 Villeneuve d'Ascq
Cedex, France}
\homepage{http://www.phlam.univ-lille1.fr/atfr/cq}
\author{Dominique Delande}
\affiliation{Laboratoire Kastler Brossel, Tour 12, Etage 1, 4 Place Jussieu, F-75005 Paris, France}
\date{\today}

\begin{abstract}

Subtle internal interference effects allow quantum-chaotic systems 
to display ``sub-Fourier" resonances, i.e. to distinguish two neighboring driving frequencies in a time shorter than the inverse of the difference 
of the two frequencies. We report experiments on the atomic version of the kicked
rotor showing the unusual 
properties of the sub-Fourier resonances, and
develop a theoretical approach (based on the Floquet theorem) 
explaining these properties, and correctly predicting the widths and lineshapes.

\end{abstract}
\maketitle

A major difference between 
classical and 
quantum systems is the
existence in the latter of \textit{interferences} between various paths. At
long times, there are typically a large number of trajectories and one could
expect that, in the average, the contributions of the various paths have
uncorrelated phases, leading to vanishingly small interference terms; in this
view, 
quantum and classical transport should be identical.
This simple expectation is however too naive. Even for classically
chaotic systems where trajectories are very complicated and proliferate
rapidly, it is not true that the phases of the various contributions are
uncorrelated. A dramatic example is the so-called dynamical localization (DL)
taking place in a periodically kicked rotor, described by the following
Hamiltonian:
\begin{equation}
H_{0}=\frac{p^{2}}{2}+\frac{K}{T}\cos\theta\sum_{n}\delta(t-nT)
\label{kicked}
\end{equation}
where $\theta$ is the $2\pi-$periodic position of the rotor, $p$ the conjugate
momentum, $T$
the period and $K$ is proportional to the strength of the kicks). 
For $K>5,$ the classical dynamics is
essentially chaotic. Although deterministic, the motion can be, on the
average, described as a diffusion in momentum space, with the average kinetic
energy growing linearly with time. The situation is completely different for
the quantum evolution: after an initial diffusive growth, the average kinetic
energy saturates 
after some break-time.

DL, leading to a complete freeze of the classical
diffusion, 
is a non-trivial effect, as it means that the quantum interferences between
classical diffusive trajectories are on the average completely destructive at
long time. A fruitful insight is obtained by using the
Floquet theorem. The so-called Floquet states $|\varphi_{k}\rangle$ are
defined as the eigenstates of the unitary evolution operator $U(T)$ associated
with $H_{0}$ over one period:
$U(T)|\varphi_{k}\rangle=\exp\left(  -i\epsilon_{k}\right)  |\varphi_{k}
\rangle$
where $\epsilon_{k}$ is the eigenphase. 
The
temporal evolution of any state $|\psi\rangle$ after $n$ periods is:
\begin{equation}
|\psi(nT)\rangle=[U(T)]^n|\psi(0)\rangle=\sum_{k}{c_{k}\ {\mathrm e}^{
-in\epsilon_{k}} \ |\varphi_{k}\rangle} \label{coef}
\end{equation}
with $c_{k}=\langle\varphi_{k}|\psi(0)\rangle.$ The evolution of any quantity
can be calculated using the Floquet basis, for example:
\begin{equation}
\langle p^{2}(nT)\rangle=\sum_{k,k^{\prime}}{c_{k}c_{k^{\prime}}^{\ast}
\ {\mathrm e}^{ -in(\epsilon_{k}-\epsilon_{k^{\prime}})}\  \langle
\varphi_{k^{\prime}}|p^{2}|\varphi_{k}\rangle.} \label{p2}
\end{equation}
The Floquet states ${|\varphi_{k}\rangle}$ of the chaotic kicked rotor are well
known: they are \emph{on the average exponentially localized} in momentum
space around a most probable momentum $p_{k},$ with a characteristic
localization length $\ell$~\cite{Casati_LocFloquetQKR_PRL90}.
Such a localization -- at the origin of DL -- is far from obvious and is closely related to the Anderson
localization in time-independent disordered one-dimensional systems~\cite{Fishman_LocDynAnderson_PRA84}.
As it is typical for classically chaotic systems, the Floquet
spectrum is highly sensitive to changes of the 
parameters, and the momentum wavefunction $\varphi_{k}(p)$ 
displays large fluctuations.

If the initial state $|\psi(0)\rangle$ is well localized in momentum space
around, say, zero momentum, Eq.~(\ref{coef}) implies that only Floquet
states with 
roughly $|p_{k}| \lesssim \ell$ 
(``important" Floquet states) will play a significant role in the dynamics. 
Eq.~(\ref{p2}) is a \emph{coherent} sum over Floquet
states. However, as times goes on, non-diagonal interference terms
accumulate larger and larger phases. In a typical chaotic system, these
phases will be uncorrelated at long times, 
and interference terms will on
the average cancel out, leading to an \emph{incoherent} sum:
\begin{equation}
\langle p^{2}\rangle\approx\sum_{k}{|c_{k}|^{2}\langle\varphi_{k}
|p^{2}|\varphi_{k}\rangle} \label{v2-DL}
\end{equation}
This 
equation is valid when DL is established. How long does
it take for the phases $n(\epsilon_{k}-\epsilon_{k^{\prime}})$ to be of the
order of $2\pi?$ This can be simply estimated from the level spacing between
important Floquet states and turns out to be roughly $t_{\mathrm{break}}=\ell
T,$ while $\langle p^{2}\rangle$ saturates to a value $\propto\ell^{2}.$

DL has been observed in experiments 
with cold atoms exposed to kicks of a time-periodic,
far detuned, standing laser wave
\cite{Raizen_LDynFirst_PRL94,Christ_LDynNoise_PRL98,AP_Bicolor_PRL00}. 
Our experimental setup is described in~\cite{AP_ChaosQTransp_CNSNS_2003}.
Basically, cold cesium atoms are produced in a standard magneto-optical trap.
The trap is turned off, and 
pulses of a far-detuned (9.2 GHz $\sim$
1700 $\Gamma)$ standing wave (around 90 mW in each direction) are applied. At the
end of the pulse series, counter-propagating phase-coherent Raman beams
perform velocity-selective Raman stimulated transitions between the 
hyperfine ground state sublevels $F_{g}=4$ and $F_{g}=3$. 
A resonant probe beam is used to
estimate the fraction of transfered atoms, thus measuring the population
of a velocity class. 
Repeated measurements 
allow to reconstruct the atomic momentum distribution $P(p)$.

\begin{figure}
\begin{center}
\includegraphics[width=4.5cm,angle=-90]{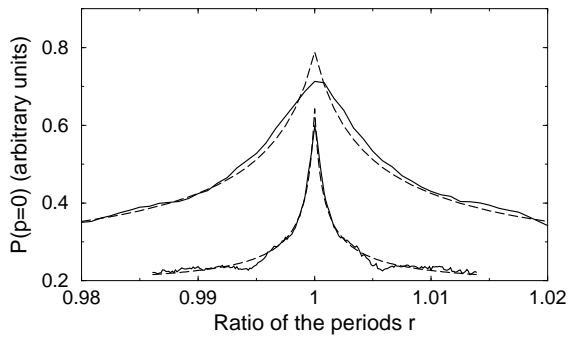}
\end{center}
\caption{
Sub-Fourier resonances. The measured atomic population in
the zero-momentum class, after 20 (upper solid curve) and 100 (lower solid
curve) double kicks is plotted vs.~the ratio $r$ of the periods.
The period of the first sequence is $T=27.8 \mu$s, 
and $K\approx 10.$ At $r=1$, dynamical localization is responsible for
the large number of zero-momentum atoms. 
Away from the exact resonance, localization is progressively destroyed. Note the narrowness 
(5 times smaller than Fourier limit) and the triangular shape of the
resonance line. Dashed lines are fits to the experimental 
lines using Eq.~(\ref{lineshape}).}
\label{pv-r}
\end{figure}

When the system is exposed to a \emph{two-frequency non-periodic driving},
no DL is expected \cite{Casati_IncommFreqsQKR_PRL89}. Consider the Hamiltonian:
\begin{equation}
H(r,\lambda)=\frac{p^{2}}{2}+\frac{K}{T}\cos\theta\sum_{n}{\left[  \delta(t-nT)+\delta
(t-nrT-\lambda T)\right]  }. \label{qp}
\end{equation}
where $r$ is the ratio of periods and $\lambda$ the initial phase
between the two kick sequences. If $r$ is rational, the system is strictly
time periodic and DL takes place, but DL is rapidly destroyed
around any rational number, as experimentally observed
in~\cite{AP_Bicolor_PRL00}. Fig.~\ref{pv-r} shows the experimentally measured
population in the zero-momentum class $P(p=0)$ after 
$20$ and $100$ double
kicks as a function 
$r.$ It displays a spectacular peak at $r=1$
and a sharp decrease on both sides, 
due to the destruction of DL. There are two surprising
features: ({\it i}) The resonance is very narrow: after $N$ kicks, it could be
argued that the two quasi-periods can be distinguished only if they differ by
$1/N$ (in relative value). This would predict a width of the order of $\Delta
r=1/N=0.01$ for 100 kicks, whereas we experimentally observe $0.0018$
\cite{AP_SubFourier_PRL02}. ({\it ii}) This ``sub-Fourier resonance 
is not smooth,
but has a marked 
cusp at the maximum. The aim of this
paper is to discuss the physical mechanism responsible for the destruction of
DL and to give an explanation for these two unexpected features.

\begin{figure}
\psfrag{p2}{$\langle p^2/\hbar^2k_{\mathrm{L}}^2\rangle $}
\par
\begin{center}
\includegraphics[width=4.5cm,angle=-90]{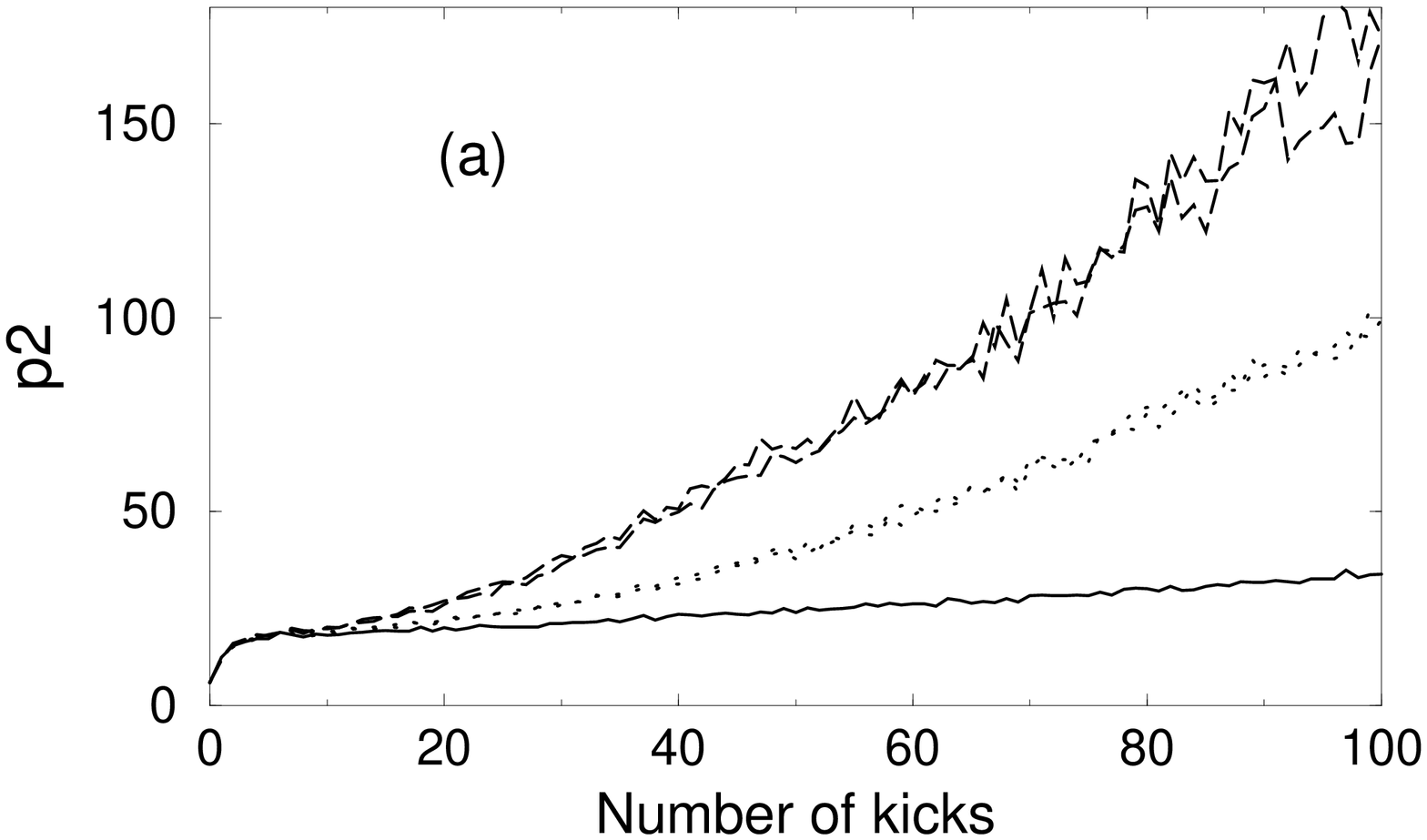}
\includegraphics[width=4.5cm,angle=-90]{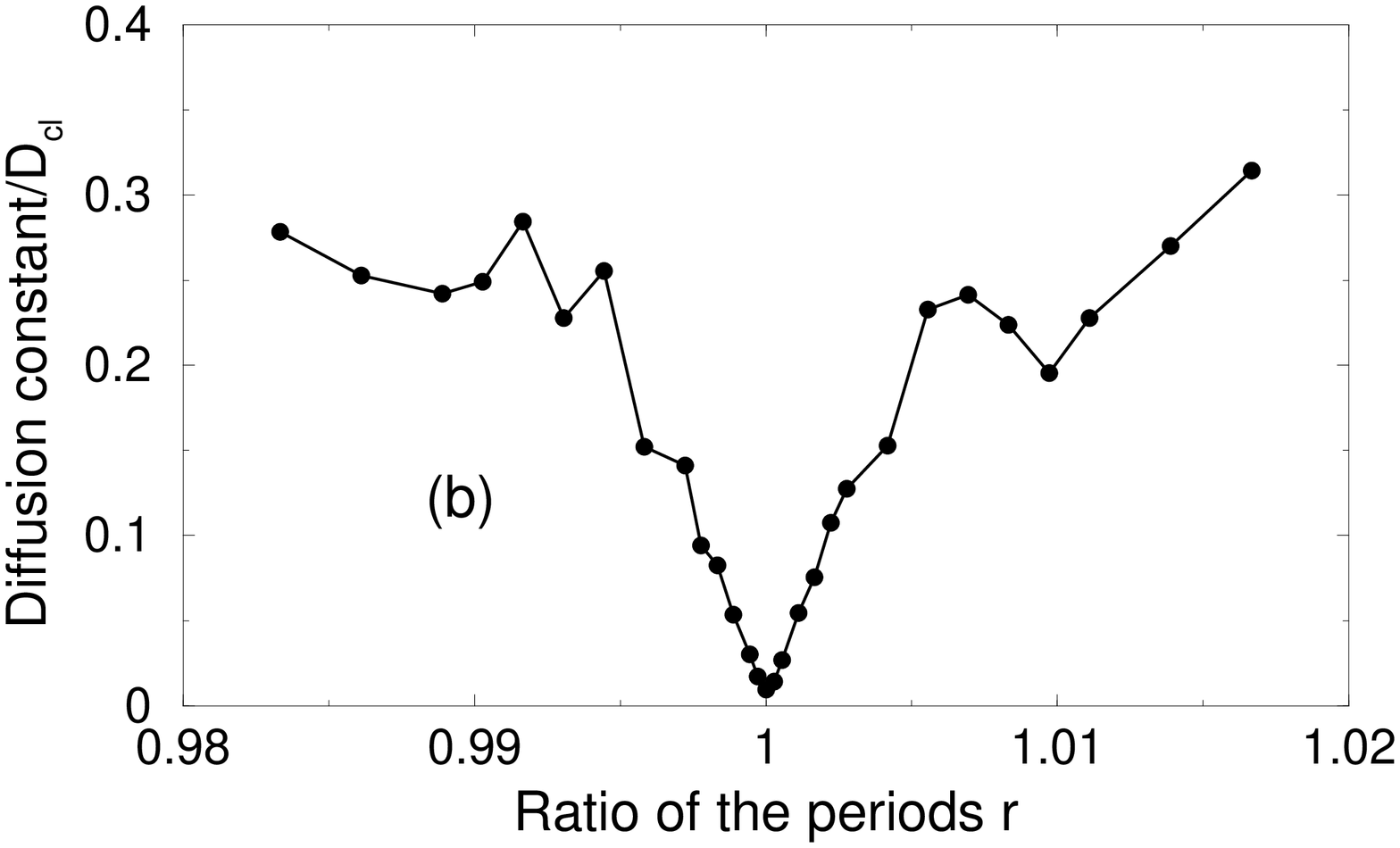}
\end{center}
\caption{(a) Averaged squared momentum $\langle p^{2} \rangle$ ($p$ is
in recoil momentum units) vs.~number of double kicks. All curves show
DL after 5-10 kicks. After this break-time, the dynamics
is frozen for the periodic system ($r$=1, solid curve), but a residual quantum
diffusion is observed in the quasi-periodic case, for $r=1+0.00111$
and $r=1-0.00111$ (dotted curves) and $r=1\pm0.00222$ (dashed curves). (b) The quantum
diffusion constant (in units of the classical diffusion constant)
versus $r,$ from data shown in (a). It (almost) vanishes at $r=1$
because of DL, and rapidly increases symmetrically on both
sides, with a triangular cusp.}
\label{diff}
\end{figure}

Let us now consider the
system evolution. Fig.~\ref{diff}(a) shows $\langle p^{2}\rangle$ as a
function of time, for various $r$ values. Note that we measure 
$P(p=0)$ and
then deduce $\langle p^{2}\rangle$ assuming that 
$\langle p^2 \rangle \propto 1/[P(p=0)]^{2}$.
For $r=1$, DL is visible after a break-time of the order of 5-10 periods, as
$\langle p^{2}\rangle$ saturates to a finite value. For the other values of $r$, the
short-time dynamics is similar,
but, for $t>t_{\mathrm{break}}$,
$\langle p^{2}\rangle$ increases slowly, linearly with time. In other words, in the
vicinity of the resonance, the quantum transport is not completely frozen, a
residual \emph{quantum} diffusive process takes place, with a diffusion
constant much smaller than the \emph{classical} constant observed at short
times. 
Fig.~\ref{diff}(b) shows the quantum diffusion constant, as a
function of $r.$ It displays a sharp minimum at $r=1$ with a characteristic
triangular shape. Due to residual spontaneous emission by the
kicked atoms \cite{Raizen_LDynNoise_PRL98}, the quantum diffusion constant at
$r=1$ is not strictly zero.

\begin{figure}
\begin{center}
\includegraphics[width=5.5cm,angle=-90]{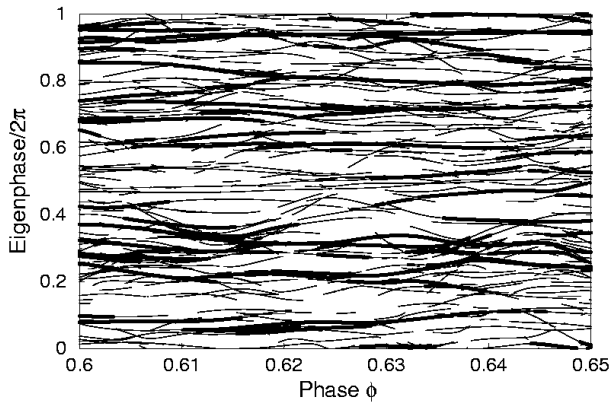}
\end{center}
\caption{Eigenphases $\epsilon_{k}$ of the evolution operator over one period
for the doubly-kicked rotor, corresponding to the experimental conditions,
versus the phase $\lambda.$ The finite duration of the 
kicks (800 ns) is
taken into account. 
We have only plotted Floquet states having a significative weight: $|\langle\psi(0)|\varphi
_{k}(\lambda_{0})\rangle|^{2}>$ $10^{-4}$ (thin line), or $>10^{-2}$ (thick
line). States appear (disappear) as their weights go above (below) the
threshold. 
Note that Floquet 
states rapidly change 
when avoided crossings with other states are encountered.}
\label{level-dynamics}
\end{figure}

The key point to understand the dynamics of the system is to realize that,
near $r=1$, the sequence of kicks is very similar to a periodic doubly-kicked
rotor for $r=1$, except that the phase $\lambda$ between the two sequences 
slowly drifts.
The evolution operator $\mathcal{U}$ of the quasi-periodic system
from time $nT$ to $(n+1)T$ is
thus given by the evolution operator $U$ of the periodic doubly-kicked rotor
for 
slowly changing $\lambda$:
$\mathcal{U}(r;\lambda
_{0};nT\rightarrow(n+1)T)=U(\lambda_{0}+n(r-1)),$ where $\lambda_{0}$
denotes the initial phase between the two pulse sequences.
The total evolution operator from $t=0$ to $(N-1)T$ can thus be written as the
product $\prod_{n=0}^{N-1}{U(\lambda_{0}+n(r-1))}.$ For 
small enough
$|r-1|,$ the \emph{adiabatic approximation} can be used,
which states that, if the system is initially prepared in a Floquet
eigenstate of $U(\lambda_{0})$ it
remains in an eigenstate of the {\em instantaneous} evolution operator $U(\lambda)$,
following the continuous deformation of the Floquet eigenstate of the periodic
doubly-kicked rotor. This is illustrated in Fig.~\ref{level-dynamics}, which
shows the Floquet spectrum of the periodic doubly-kicked rotor, obtained by
numerically diagonalizing 
$U(\lambda)$.
When 
$\lambda$ is varied, the energy levels evolve in a complex manner.
The complicated ``spaghetti" behaviour, characteristic of
quantum-chaotic systems, with a large number of avoided crossings (ACs) is
directly responsible for the sub-Fourier character of the resonance lines,
as we show below.

In our case, the initial state is 
a linear combination of Floquet eigenstates. In the adiabatic approximation, it remains
a linear combination of the instantaneous Floquet states with 
the same
weights (the phases change, but the squared moduli remain constant). As
discussed above, the coherences between Floquet states play a role only at
short time, and interference terms average to zero after the break-time,
Eq.~(\ref{v2-DL}). 
This
implies that Eq.~(\ref{v2-DL}) remains valid close to the resonance, provided
one uses the instantaneous Floquet eigenbasis:
\begin{equation}
\langle p^{2}(\lambda)\rangle\approx\sum_{k}{|\langle\psi(0)|\varphi
_{k}(\lambda_{0})\rangle|^{2}\ \langle\varphi_{k}(\lambda)|p^{2}|\varphi
_{k}(\lambda)\rangle.} \label{v2-new}
\end{equation}
The weights $|\langle\psi(0)|\varphi_{k}(\lambda_{0})\rangle|^{2}$ are
estimated at the \emph{initial} time
($\lambda=\lambda_{0},$), while the
average values $\langle\varphi_{k}(\lambda)|p^{2}|\varphi_{k}(\lambda)\rangle$
are 
calculated at the \emph{final} value $\lambda=\lambda
_{0}+N(r-1).$ Eq.~(\ref{v2-new}) explains the experimental observations. The
fundamental 
point is that Eq.~(\ref{v2-new}) describes a correlation between
the localization properties of the Floquet eigenstates for two values of
$\lambda.$ The $|\langle\psi(0)|\varphi_{k}(\lambda_{0})\rangle|^{2}$ term
gives important weights to the Floquet states localized near zero momentum.
As $\lambda$ moves away from $\lambda_{0}$, the important Floquet states
evolve and, on the average, their center moves away from zero momentum. Thus,
one expects 
$\langle p^{2}(\lambda)\rangle$ to have a minimum at $\lambda
=\lambda_{0}$ and to rapidly increase on both sides of $\lambda_{0}.$ Since
$\lambda-\lambda_{0}$ is proportional to $r-1$, 
$\langle p^{2}\rangle$ shall present a sharp
minimum at $r=1;$ simultaneously, the population in the zero-momentum class
shall present a sharp maximum at $r=1,$ 
which is experimentally observed in Fig.~\ref{pv-r}.

What 
determines the linewidth of the resonances?
A state $|\varphi
_{k}(\lambda)\rangle$ evolves because of its ACs with other Floquet eigenstates.
Tiny ACs may be crossed diabatically,
leading to a large modification of the state. On the average,
the scale $\Delta\lambda=\lambda-\lambda_{0}$ on which $|\varphi_{k}
(\lambda)\rangle$ loses the localization property of $|\varphi_{k}(\lambda
_{0})\rangle$ is the distance to the next AC. We immediately deduce that
the width $\Delta f$ of the sub-Fourier line is such that:
\begin{equation}
NT\Delta f=2\Delta\lambda.\label{width}
\end{equation}
Eq.~(\ref{width}) also explains why chaos is necessary to observe sub-Fourier
resonances. If the classical dynamics is regular, the Floquet eigenstates
evolve smoothly with the parameter $\lambda$;
a change in
$\lambda$ of the order of one is thus required
to significantly modify the Floquet states:
$2\Delta\lambda\approx 1$, which is 
the Fourier limit. On the
contrary, in a classically chaotic system, the level dynamics is known to be
complex with plenty of ACs (Fig.~\ref{level-dynamics}), $\Delta\lambda$ is
then much smaller than unity, leading to highly sub-Fourier resonances (up to
a factor 37 observed in \cite{AP_SubFourier_PRL02}). Eq.~(\ref{width}) also predicts
the linewidth 
to be inversely proportional to the temporal length of
the kick sequence
beyond the break-time  (i.e. the sub-Fourier character is independent of $N$), 
as numericaly observed \cite{AP_SubFourier_PRL02}. The
value of $\Delta\lambda$ depends on the detailed dynamics of the system. It
can be roughly estimated by visual inspection of the quasi-energy level
dynamics, Fig.~\ref{level-dynamics}. The level spacing between important
Floquet states is the inverse of the number of such states, itself roughly
equal to the localization length, of the order of 5 for the present set of
parameters. The typical 
{\em relative} slope of neighbor levels in Fig.~\ref{level-dynamics} 
is of the order of
4, thus predicting $\Delta\lambda$ of the order of 0.05 and a
``factor 10" sub-Fourier line, about twice the
experimentally observed factor 
(the difference is due to the transverse profile
of the laser mode leading to spatial inhomogeneities in $K$).

From Eq.~(\ref{v2-new}) and the previous analysis,
it is expected that $\langle p^{2}
\rangle$ 
significantly increases from its minimum value
at $\lambda=\lambda_{0}$ 
as $\lambda$ varies to $\lambda_0+\Delta\lambda.$ It seems
reasonable to assume that this increase is initially quadratic in
$\lambda-\lambda_{0}$ \cite{note:RMT}.
The
situation here is more complicated because Floquet 
states are localized in momentum space. ACs between Floquet states localized 
around the same 
momentum, will be rather large 
(this is what determines 
$\Delta\lambda$)
whereas ACs between states 
localized a distance 
$L \gg\ell$ apart in momentum space are typically much smaller and scale like $\exp(-L/\ell).$
There is thus a very broad distribution of AC widths, with a
large number of tiny ACs. A tiny AC typically extends over a small $\lambda$
interval and thus tends to produce small values of $\Delta\lambda.$ 
The increase of
$\langle p^{2}\rangle$ 
with $\lambda$
thus depends on the number of small ACs encountered. In
the presence of exponential localization, the 
AC density
scales with size $C$ as $1/C$ for $C\rightarrow 0$, 
and $\langle p^{2}(\lambda)\rangle-\langle p^{2}(\lambda_{0})\rangle$ shall behave like
$|\lambda-\lambda_{0}|$ \cite{note:ACdensity}, producing the cusp
experimentally observed in the resonance line, Fig.~\ref{pv-r}, and in the
diffusion constant, Fig.~\ref{diff}. The  
large number of
extremely small $C$ is responsible for the singularity of the sub-Fourier
resonance line. Another consequence is the diffusive behaviour observed in the
vicinity of the resonance, see Fig.~\ref{diff}(b). Indeed, $\langle p^{2}
(\lambda)\rangle-\langle p^{2}(\lambda_{0})\rangle$ increases linearly with
$|\lambda-\lambda_{0}|,$ itself a linear function of time and $|r-1|.$ Thus,
our 
model correctly predicts two non-trivial properties : $\langle
p^{2}\rangle$ increases linearly with time and the corresponding diffusion
constant is proportional to $|r-1|.$

There will always be some degree of nonadiabaticity. Whatever small $|r-1|$
is, 
tiny enough ACs will be crossed 
diabatically. This puts a lower bound on the size of the ACs effectively
participating in the quantum dynamics of the system and produces a
rounding of the top of the sub-Fourier resonance line, 
which is too
small to be seen in the experiment after 100 kicks, but is easily visible after 20
kicks, Fig.~\ref{pv-r}.

Our approach concentrates on the immediate vicinity of
the resonance. What happens  
in the wings of the sub-Fourier line?
Eq.~(\ref{v2-new}) 
indicates that this depends on the residual
correlation between $|\varphi_{k}(\lambda_{0})\rangle$ and $|\varphi
_{k}(\lambda)\rangle$ 
for $|\lambda-\lambda_{0}|>\Delta\lambda.$
A quantitative answer to this question
cannot be given, as there is no typical AC width. 
However, it seems clear that the quantum diffusion
constant 
does not exceed the classical one,
which correspons to the killing of all interference terms. 
Random Matrix Theory tells us that this type of
parametric correlation usually decays 
algebraically with $\lambda-\lambda_{0}.$ We thus propose the following 
{\em anzatz}:
\begin{equation}
\langle p^{2}(nT)\rangle=\langle p^{2}\rangle_{DL}+D_{cl}\frac
{|r-1|}{|r-1|+\Delta\lambda/t_{\mathrm{break}}} nT
\label{lineshape}
\end{equation}
where $D_{cl}$ is the classical diffusion constant and $\langle p^{2}
\rangle_{DL}$ the 
saturation
value of $p^{2}$ 
due to DL. This equation fits 
(using $P(p=0)\propto \langle p^2 \rangle ^{-1/2}$)
very well the experimental curves 
in Fig.~\ref{pv-r},
reproducing the
linear behaviour at the center of the resonance and the
classical diffusion 
in the wings.
Nevertheless, because of the laser inhomogeneities, the parameters are fitted, 
not extracted from the previous analysis.

In summary, we have developed a theoretical approach for the mechanism of
sub-Fourier resonances, which correctly predicts the unexpected 
observed features.
Deviations from exact periodicity are treated in the
framework of the adiabatic approximation for the Floquet spectrum of the
system, which thus goes beyond a pertubative approach. 
The dynamics is governed by instantaneous Floquet eigenstates,
which are non-trivial objects, as they need to have very well defined internal (i.e. between the
various parts of the wavefunction for each Floquet state) phase coherence to
be stationary states of the strictly periodic system, but at the same time,
the inter-state coherences do not play any role (beyond the break-time,
Floquet states effectively add incoherently). We are not aware of any other
quantum system where the dynamics is dominated by such an \emph{incoherent}
sum of internally extremely \emph{coherent} states. It shows that the role of
interferences in quantum mechanics is far from obvious, and can produce unexpected
behaviors.

Laboratoire de Physique des
Lasers, Atomes et Molécules (PhLAM) is Unité Mixte de Recherche
UMR 8523 du CNRS et de l'Université des Sciences et Technologies de
Lille.
Laboratoire Kastler Brossel is laboratoire 
de l'Universit{\'e} Pierre et Marie
Curie et de l'Ecole Normale Sup{\'e}rieure, UMR 8552 du CNRS. CPU time on various computers has been provided by IDRIS.


\begin{thebibliography}{11}
\expandafter\ifx\csname natexlab\endcsname\relax\def\natexlab#1{#1}\fi
\expandafter\ifx\csname bibnamefont\endcsname\relax
  \def\bibnamefont#1{#1}\fi
\expandafter\ifx\csname bibfnamefont\endcsname\relax
  \def\bibfnamefont#1{#1}\fi
\expandafter\ifx\csname citenamefont\endcsname\relax
  \def\citenamefont#1{#1}\fi
\expandafter\ifx\csname url\endcsname\relax
  \def\url#1{\texttt{#1}}\fi
\expandafter\ifx\csname urlprefix\endcsname\relax\def\urlprefix{URL }\fi
\providecommand{\bibinfo}[2]{#2}
\providecommand{\eprint}[2][]{\url{#2}}

\bibitem[{\citenamefont{Casati et~al.}(1990)\citenamefont{Casati, Guarneri,
  Izrailev, and Scharf}}]{Casati_LocFloquetQKR_PRL90}
\bibinfo{author}{\bibfnamefont{G.}~\bibnamefont{Casati}},
  \bibinfo{author}{\bibfnamefont{I.}~\bibnamefont{Guarneri}},
  \bibinfo{author}{\bibfnamefont{F.~M.} \bibnamefont{Izrailev}},
  \bibnamefont{and} \bibinfo{author}{\bibfnamefont{R.}~\bibnamefont{Scharf}},
  \bibinfo{journal}{Phys. Rev. Lett.} \textbf{\bibinfo{volume}{64}},
  \bibinfo{pages}{5} (\bibinfo{year}{1990}).

\bibitem[{\citenamefont{Grempel et~al.}(1984)\citenamefont{Grempel, Prange, and
  Fishman}}]{Fishman_LocDynAnderson_PRA84}
\bibinfo{author}{\bibfnamefont{D.~R.} \bibnamefont{Grempel}},
  \bibinfo{author}{\bibfnamefont{R.~E.} \bibnamefont{Prange}},
  \bibnamefont{and} \bibinfo{author}{\bibfnamefont{S.}~\bibnamefont{Fishman}},
  \bibinfo{journal}{Phys. Rev. A} \textbf{\bibinfo{volume}{29}},
  \bibinfo{pages}{1639} (\bibinfo{year}{1984}).

\bibitem[{\citenamefont{Moore et~al.}(1994)\citenamefont{Moore, Robinson,
  Bharucha, Williams, and Raizen}}]{Raizen_LDynFirst_PRL94}
\bibinfo{author}{\bibfnamefont{F.~L.} \bibnamefont{Moore}},
  \bibinfo{author}{\bibfnamefont{J.~C.} \bibnamefont{Robinson}},
  \bibinfo{author}{\bibfnamefont{C.~F.} \bibnamefont{Bharucha}},
  \bibinfo{author}{\bibfnamefont{P.~E.} \bibnamefont{Williams}},
  \bibnamefont{and} \bibinfo{author}{\bibfnamefont{M.~G.}
  \bibnamefont{Raizen}}, \bibinfo{journal}{Phys. Rev. Lett.}
  \textbf{\bibinfo{volume}{73}}, \bibinfo{pages}{2974} (\bibinfo{year}{1994}).

\bibitem[{\citenamefont{Ammann et~al.}(1998)\citenamefont{Ammann, Gray,
  Shvarchuck, and Christensen}}]{Christ_LDynNoise_PRL98}
\bibinfo{author}{\bibfnamefont{H.}~\bibnamefont{Ammann}},
  \bibinfo{author}{\bibfnamefont{R.}~\bibnamefont{Gray}},
  \bibinfo{author}{\bibfnamefont{I.}~\bibnamefont{Shvarchuck}},
  \bibnamefont{and}
  \bibinfo{author}{\bibfnamefont{N.}~\bibnamefont{Christensen}},
  \bibinfo{journal}{Phys. Rev. Lett.} \textbf{\bibinfo{volume}{80}},
  \bibinfo{pages}{4111} (\bibinfo{year}{1998}).

\bibitem[{\citenamefont{Ringot et~al.}(2000)\citenamefont{Ringot, Szriftgiser,
  Garreau, and Delande}}]{AP_Bicolor_PRL00}
\bibinfo{author}{\bibfnamefont{J.}~\bibnamefont{Ringot}},
  \bibinfo{author}{\bibfnamefont{P.}~\bibnamefont{Szriftgiser}},
  \bibinfo{author}{\bibfnamefont{J.~C.} \bibnamefont{Garreau}},
  \bibnamefont{and} \bibinfo{author}{\bibfnamefont{D.}~\bibnamefont{Delande}},
  \bibinfo{journal}{Phys. Rev. Lett.} \textbf{\bibinfo{volume}{85}},
  \bibinfo{pages}{2741} (\bibinfo{year}{2000}).

\bibitem[{\citenamefont{Szriftgiser et~al.}(2003)\citenamefont{Szriftgiser,
  Lignier, Ringot, Garreau, and Delande}}]{AP_ChaosQTransp_CNSNS_2003}
\bibinfo{author}{\bibfnamefont{P.}~\bibnamefont{Szriftgiser}},
  \bibinfo{author}{\bibfnamefont{H.}~\bibnamefont{Lignier}},
  \bibinfo{author}{\bibfnamefont{J.}~\bibnamefont{Ringot}},
  \bibinfo{author}{\bibfnamefont{J.~C.} \bibnamefont{Garreau}},
  \bibnamefont{and} \bibinfo{author}{\bibfnamefont{D.}~\bibnamefont{Delande}},
  \bibinfo{journal}{Commun. Nonlin. Sci. Num. Simul.}
  \textbf{\bibinfo{volume}{8}}, \bibinfo{pages}{301} (\bibinfo{year}{2003}).

\bibitem[{\citenamefont{Casati et~al.}(1989)\citenamefont{Casati, Guarnieri,
  and Shepelyansk}}]{Casati_IncommFreqsQKR_PRL89}
\bibinfo{author}{\bibfnamefont{G.}~\bibnamefont{Casati}},
  \bibinfo{author}{\bibfnamefont{I.}~\bibnamefont{Guarnieri}},
  \bibnamefont{and} \bibinfo{author}{\bibfnamefont{D.~L.}
  \bibnamefont{Shepelyansk}}, \bibinfo{journal}{Phys. Rev. Lett.}
  \textbf{\bibinfo{volume}{62}}, \bibinfo{pages}{345} (\bibinfo{year}{1989}).

\bibitem[{\citenamefont{Szriftgiser et~al.}(2002)\citenamefont{Szriftgiser,
  Ringot, Delande, and Garreau}}]{AP_SubFourier_PRL02}
\bibinfo{author}{\bibfnamefont{P.}~\bibnamefont{Szriftgiser}},
  \bibinfo{author}{\bibfnamefont{J.}~\bibnamefont{Ringot}},
  \bibinfo{author}{\bibfnamefont{D.}~\bibnamefont{Delande}}, \bibnamefont{and}
  \bibinfo{author}{\bibfnamefont{J.~C.} \bibnamefont{Garreau}},
  \bibinfo{journal}{Phys. Rev. Lett.} \textbf{\bibinfo{volume}{89}},
  \bibinfo{pages}{224101} (\bibinfo{year}{2002}).

\bibitem[{\citenamefont{Klappauf et~al.}(1998)\citenamefont{Klappauf, Oskay,
  Steck, and Raizen}}]{Raizen_LDynNoise_PRL98}
\bibinfo{author}{\bibfnamefont{B.~G.} \bibnamefont{Klappauf}},
  \bibinfo{author}{\bibfnamefont{W.~H.} \bibnamefont{Oskay}},
  \bibinfo{author}{\bibfnamefont{D.~A.} \bibnamefont{Steck}}, \bibnamefont{and}
  \bibinfo{author}{\bibfnamefont{M.~G.} \bibnamefont{Raizen}},
  \bibinfo{journal}{Phys. Rev. Lett.} \textbf{\bibinfo{volume}{81}},
  \bibinfo{pages}{1203} (\bibinfo{year}{1998}).

\bibitem[{not({\natexlab{a}})}]{note:RMT}
\bibinfo{note}{A Random Matrix Theory model indeed predicts such a behaviour:
  B. D. Simons and B. L. Altshuler, Phys. Rev. Lett. {\bf 70}, 4063 (1993).}

\bibitem[{not({\natexlab{b}})}]{note:ACdensity}
\bibinfo{note}{{This basically comes from the fact that the probability to
  encounter a ``bad" AC which expels the atoms from the zero-momentum region is
  proportional to the length of the interval $|\lambda-\lambda_{0}|$}}.

\end{thebibliography}

\end{document}